\documentclass[12pt,preprint]{aastex}

\slugcomment{Not to appear in Nonlearned J., 45.}

\shorttitle{The local universe as seen in FIR and in FUV}
\shortauthors{Buat et al.}

\begin{document}


\title{The local universe as seen in far-infrared and in far-ultraviolet: a global point of view on the local recent star formation}


\author{V. Buat\altaffilmark{1}, 
T.\ T.\ Takeuchi \altaffilmark{1,2}, 
J. Iglesias-P\'{a}ramo \altaffilmark{3}, 
C.\ K.\ Xu\altaffilmark {4}, 
D.\ Burgarella\altaffilmark{1},
A. Boselli\altaffilmark{1},
T.\ Barlow\altaffilmark{2}, 
L.\ Bianchi \altaffilmark{5}, 
J.\ Donas\altaffilmark{1}, 
K.\ Forster\altaffilmark{4}, 
P.\ G.\ Friedman\altaffilmark{4}, 
T.\ M.\ Heckman\altaffilmark{7},  
Y.\ -W.\ Lee\altaffilmark{6}, 
B.\ F.\ Madore\altaffilmark{10,11}, , 
D.\ C.\ Martin\altaffilmark{4}, 
B.\ Milliard\altaffilmark{1}, 
P.\ Morissey\altaffilmark{4}, 
S.\ Neff\altaffilmark{12}, 
M.\ Rich\altaffilmark{9}, 
D.\ Schiminovich\altaffilmark{13},
M.\ Seibert\ \altaffilmark{4},
T.\ Small\altaffilmark{4}, 
A.\ S.\ Szalay\altaffilmark{7}, 
B.\ Welsh\altaffilmark{8} 
T.\ Wyder\altaffilmark{4}
and
S. K.\ Yi\altaffilmark{6}}

\altaffiltext{1}{
Laboratoire d'Astrophysique de Marseille, Marseille 13012, France }
\altaffiltext{2}{
Astronomical Institute, Tohoku University, Aoba, Aramaki, Aoba-ku, Sendai 
980--8578, Japan}
\altaffiltext{3}{
Instituto de Astrof\'{\i}sica de Andaluc\'{\i}a,CSIC, 18008 Granada, SPAIN}
\altaffiltext{4}{
California Institute of Technology, MC 405-47, 1200 East California 
Boulevard, Pasadena, CA 91125}
\altaffiltext{5}{
Center for Astrophysical Sciences, The Johns Hopkins University, 3400 
N.\ Charles St., Baltimore, MD 21218}
\altaffiltext{6}{
Center for Space Astrophysics, Yonsei University, Seoul 120--749, Korea}
\altaffiltext{7}{
Department of Physics and Astronomy, The Johns Hopkins
University, Homewood Campus, Baltimore, MD 21218}
\altaffiltext{8}{
Space Sciences Laboratory, University of California at
Berkeley, Berkeley, CA 94720}
\altaffiltext{9}{
Department of Physics and Astronomy, University of California, Los Angeles, 
CA 90095}
\altaffiltext{10}{
Observatories of the Carnegie Institution of Washington, 813 Santa Barbara 
St., Pasadena, CA 91101}
\altaffiltext{11}{
NASA/IPAC Extragalactic Database, California Institute
of Technology, Mail Code 100-22, 770 S. Wilson Ave., Pasadena, CA 91125}
\altaffiltext{12}{
Laboratory for Astronomy and Solar Physics, NASA Goddard Space Flight Center, 
Greenbelt, MD 20771}
\altaffiltext{13}{
Department of Astronomy, Columbia University, New York, NY 10027}


\begin{abstract}
We select far-infrared (FIR-$60\;\mu$m) and far-ultraviolet (FUV-1530~\AA) samples 
of nearby galaxies in order to discuss the biases encountered by 
monochromatic surveys (FIR or FUV). 
Very different volumes are sampled by each selection and much care is 
taken to apply volume corrections to all the analyses. 
The distributions of the bolometric luminosity of young stars are compared 
for both samples: they are found to be consistent with each other for 
galaxies of intermediate luminosities but some differences are found for high 
($>5 ~ 10^{10} L_\sun$) luminosities. The shallowness of the {\sl IRAS} survey 
prevents us from  securing comparison at low luminosities 
($<2 ~ 10^9 L_\sun$).
The ratio of the total infrared (TIR) luminosity to the FUV luminosity is found to increase 
with the bolometric luminosity in a similar way for both samples up to 
$5 ~ 10^{10} L_\sun$. 
Brighter galaxies are found to have a different behavior according to 
their selection: the $L_{\rm TIR}/L_{FUV}$ ratio of the FUV-selected galaxies 
brighter than $5~10^{10} L_\sun$ reaches a plateau whereas 
$L_{\rm TIR}/L_{FUV}$ 
continues to increase with the luminosity of bright galaxies selected in FIR. 
The volume-averaged specific star formation rate (SFR per unit galaxy stellar 
mass, SSFR) is found to decrease toward massive galaxies within each 
selection. 
The SSFR is found to be larger than that measured for optical and NIR-selected 
sample over the whole mass range for the FIR selection,  and for masses 
larger than $10^{10}\; M_\sun$ for the FUV selection. 
Luminous and massive galaxies selected in FIR appear as active as galaxies 
with similar characteristics detected at $z\sim 0.7$.
\end{abstract}


\section{Introduction}

Many of the most recent galaxy surveys have attempted to gain a better
understanding of the evolution of the star formation rate (SFR) with
time and environment.  Because of the spectral redshifting, deep (high
redshift) optical surveys in fact sample the far-ultraviolet (FUV)
rest-frame emission of the target galaxies.  As a consequence,
numerous measurements of the star formation activity of galaxies as a
function of redshift ($z$) are based on restframe FUV data obtained
from imaging and spectroscopic surveys: at low z
\citep[e.g.,][]{lilly,schim,baldry} or at higher z, using drop-out
selection techniques
\citep[e.g.,][]{steidel,bunker,ouchi,giavalisco}. However, the
attenuation of the FUV light by interstellar dust is a major issue in
the derivation of quantitative SFR from the FUV even at low $z$
\citep[e.g.,][]{buat05,seibert,cortese}.

Recent infrared surveys \citep[e.g.,][]{flores,lefloch,perez} have
also contributed significantly to the study of the star formation
history in the universe
\citep[e.g.,][]{rowan01,takeuchi01a,takeuchi01b,lagache03}: the
far-infrared (FIR) emission from the dust heated by hot stars is, by
definition, not affected by dust attenuation. However, the FIR
emission has its own drawbacks: the calibration of the dust emission
into a quantitative SFR usually relies on the strong assumption that
most of the dust heating is due to young stars and that all the light
from these stars is absorbed by dust and re-emitted in FIR
\citep[e.g.,][]{kennicutt98}.  This is clearly an over-simplification:
most galaxies are seen to emit in the FUV-optical and, as a
consequence all the light emitted by young stars cannot have been
absorbed by dust.  Moreover, substantial dust heating by older stars
cannot be ruled out for all galaxies and more complex calibrations
have to be undertaken \citep{buatxu,bell03,hirashita,igle06}.  Another
issue is that infrared observations are carried out at one or a few
wavelengths (in the mid and/or far-infrared) whereas it is the total
infrared (TIR) emission that is required for a star formation rate
calculation. Unfortunately, the TIR emission is derived by
extrapolation from measured only those few measured fluxes using
models, and the correction factors range from 2 to 3, when observations
are made in the far infrared, and up to 8-10 when only mid-infrared
data are available \citep[e.g.,][]{ttt3,lefloch}.

The  ultraviolet and  the FIR  emission  are each  strongly linked  to
alternate  manifestations  of  the  recent star  formation  rate:  the
''transparent" one in FUV and the ''hidden" one in FIR. Obviously a very
promising  way to  proceed would  be  to combine  both wavelengths  to
perform a  more inclusive and multiwavelength analysis  of the current
star formation in the universe.

What do we know today about the FUV and FIR universe?  From a global
 point of view, the recent observations conducted by {\sl SPITZER} and
 {\sl GALEX} have provided insight into the TIR and FUV luminosity
 functions and densities from $z=0$ to $z=1$
 \citep[][]{schim,lefloch,perez,arn}. The shapes of the luminosity
 functions in both wavelength ranges are found to be very different
 \citep[][]{ttt2} as previously emphasized by \citet{buat98}.  The
 evolution of the luminosity functions and the derived star formation
 densities were studied in the FUV \citep[][]{schim} and at infrared
 wavelengths \citep[][]{lefloch,perez}: a strong evolution was found
 at both wavelengths, with a net increase of the luminosity density
 from $z=0$ to $z=1$. Nevertheless the evolution appears to be
 stronger in the FIR than in the FUV implying a {\it global} increase
 of the dust attenuation from $z=0$ to $z=1$ by $\sim 1\;\mbox{mag}$
 \citep[][]{ttt2}. Such an increase might be explained, at least
 qualitatively, by the larger fraction of bright galaxies at high
 redshift, combined with the known positive correlation of the dust
 attenuation with the absolute luminosity of the host galaxies
 \citep[e.g.][]{wang,buat98,hopkins01,martin}. In a recent analysis of
 {\sl SWIRE/GALEX} data \citet{xu06apjs} find no significant
 differences between the FIR-to-FUV flux ratios of star-forming galaxies
 at z=0.6 and their local counterparts of similar SFR. They argue that
 the evolution of the dust attenuation with redshift is primarily due
 to the SFR evolution and to the strong dependence of the attenuation
 on SFR itself.

Indeed, most recent studies based on large surveys aim at better understanding of which galaxies are at the 
origin of the variation of the star formation density with $z$, and especially its decrease from 
$z\sim 1$ to 0, which is seen at almost all wavelengths \citep[e.g.,][and references therein]{hopkins06}. 
{}From the UV-optical side, almost all recent surveys have found a strong increase in 
luminosity and/or space density of late-type, blue galaxies although some discrepancies have been noted.
Differences in the definition of galaxy types from one a study to another makes a direct 
comparison of the results problematic \citep[e.g.,][and references therein]{delapparent,wolf}. 
{}From the IR side, recent results from the {\sl SPITZER} Space Telescope seem to attribute the 
general decrease of the star formation density to a decrease in the SFR in massive spirals 
\citep[][]{bell} without strong interactions. Studies, both in the IR and in the optical, suggest a 
minimal role of strong mergers in the evolution of the star formation density from z=1 to z=0 
\citep[][]{lotz,bell}

{}Connecting from low- to high-$z$ that which is seen in the
rest-frame FIR or in FUV, is a new challenge.  Do we see the same
galaxy populations in FUV and FIR evolving as a whole and appearing
differently in FUV and in FIR, or must we invoke sub-populations of
galaxies evolving independently with $z$ and/or being detected only at
one wavelength, FUV or FIR? This is a crucial question: if we are
observing the same populations in both wavelengths ranges, then, with
some justification, we can try to predict the total star formation
from single-band surveys (assuming some relevant corrections). But if
other populations are present, it would appear to be impossible to
reconstruct all of the star formation activity from a single-band
survey.

The first step consists of obtaining a reliable reference dataset in
the local universe.  Thanks to {\sl GALEX}, large samples of nearby
galaxies observed in the FUV are now available. Making use of the
existing IR surveys, we can then build robust reference samples of galaxies
selected in the FUV and FIR  which are suitable for
comparison.  The aim of this paper is to take advantage of the {\sl
GALEX} shallow survey to build large samples of nearby galaxies
selected in the FUV (or in the FIR by {\sl IRAS}) and to use these samples to
analyse the selection biases and the consistency of the FIR and FUV
LFs at $z=0$.  We can then build large reference samples of galaxies
selected  at $60\;\mu$m, say,  in such a way as to allow a
very good detection rate at 1530\AA\, and vice versa (section 2).
In section 3 we emphasize the intrinsic differences existing
between a FIR and a FUV selection.  The relative contributions of the
TIR and FUV emissions to the luminosity of the young stellar
populations in galaxies is assessed in section 4 using a
cross-comparison of the luminosity functions in both samples;
bolometric luminosity functions are then built to check if we can indeed see all
the galaxies at a single wavelength. In section 5, we discuss the
star formation activity as a function of the stellar mass in both
samples through an analysis of the specific star formation rate (SSFR,
SFR per unit galaxy stellar mass). Section 6 is devoted to the
conclusions. This works extends the earlier studies of \citet{martin},
\citet{igle06} and \citet{xu06}, which were based on smaller samples.

Throughout this article, we use the cosmological concordance parameters of $H_0 = 72$ km s$^{-1}$ Mpc$^{-1}$, 
$\Omega_M = 0.3$ and $\Omega_{\lambda} = 0.7$. All magnitudes are quoted in the AB system.

\section{The galaxy samples}

\subsection{The FIR-selected sample} 

The  FIR-selected  sample  was  selected from  the  {\sl IRAS}  PSC$z$
\citep{saunders00}.   We  selected   all  of  the  confirmed  galaxies
(reliability  $\ge~50\%$ in the  PSC$z$) in  the $\sim$  3,000 deg$^2$
covered by the first public release of the {\sl GALEX} All Sky Imaging
Survey  (AIS): 777  sources not  contaminated by  cirrus  (cirrus flag
lower  than  2)  were  retrieved  over  an  effective  area  of  2,210
$\mbox{deg}^2$.   We  then extracted  FUV  images  for  each of  these
sources:  most of these  objects are  resolved by  {\sl GALEX}  and we
performed  the photometry  manually for  each source,  since  the {\sl
GALEX} pipeline  reduction currently works only for  point sources.  A
detailed  description   of  the   photometric  process  is   given  in
\citet{igle06}.  77 galaxies were not detected in the FUV; 28 of these
77 non-detected sources were located  near the edge of the {\sl GALEX}
field  where  the  image   quality  is  degraded.  For  the  remaining
non-detections, we  adopted an  upper limit to  $\mbox{FUV} =  20.5 \;
\mbox{mag}$ corresponding to the  3$\sigma$ detection limit in the AIS survey
\citep[][]{morissey}.  At  this point, we  were left with  749 galaxies
from the PSC$z$ which have either an FUV measurement or an upper limit.

The lower spatial resolution of {\sl IRAS} observations sometimes leads  to 
confusion in the selection of the FUV counterpart for a given FIR source. 
In order to check this we searched for neighbours detected by 2MASS or NVSS 
within a radius of 1~arcmin around the {\sl IRAS} position. 
We checked individually all {\sl IRAS} sources with several neighbours by 
superimposing the 2MASS, {\sl GALEX}, {\sl IRAS} and NVSS images. 
Out of the 749 sources, 63 are considered as confused.
These galaxies are  kept for the determination of the FIR LF, but they are not 
included in the analysis of the FIR and FUV properties of individual galaxies.

In the end, we are left with a sample of 686 unconfused sources for
which an FUV detection (or an upper limit) is available.  21 galaxies
are noted in the NASA/IPAC Extragalactic Database (NED), as being
active galaxies.  This number must be taken as a lower limit since
such detailed information in NED is available for only $\sim$ 2/3 of
the sample galaxies. Nevertheless, the contamination of our sample by
active galaxies is here estimated to be lower than 1.5$\%$.  In order
to have reliable distances (as determined from expansion velocities)
we consider only the 665 galaxies with velocities $v> 1,000
~\mbox{km\,s}^{-1}$.  The final sample of 665 sources is presented in
Table 1: the FUV fluxes are corrected for foreground Galactic
extinction using the \citet[][]{schlegel} dust map and the
\citet[][]{cardelli} extinction curve.  77 of these sources were not
detected at 100 $\mu$m; and for these galaxies we estimated the fluxes
at $100\;\mu$m using a mean value of $F_{60}/F_{100}=0.5
~(\sigma=0.2)$ derived from the FIR galaxies in our sample that were
detected at both wavelengths.  {}From the 2MASS survey we added in
$H$-band data: 621 out of 665 galaxies (i.e., 93~\%) have an $H$-band
magnitude.

\subsection{ The FUV-selected sample} 

\begin{figure}
 \includegraphics [angle=-90,width=18cm]{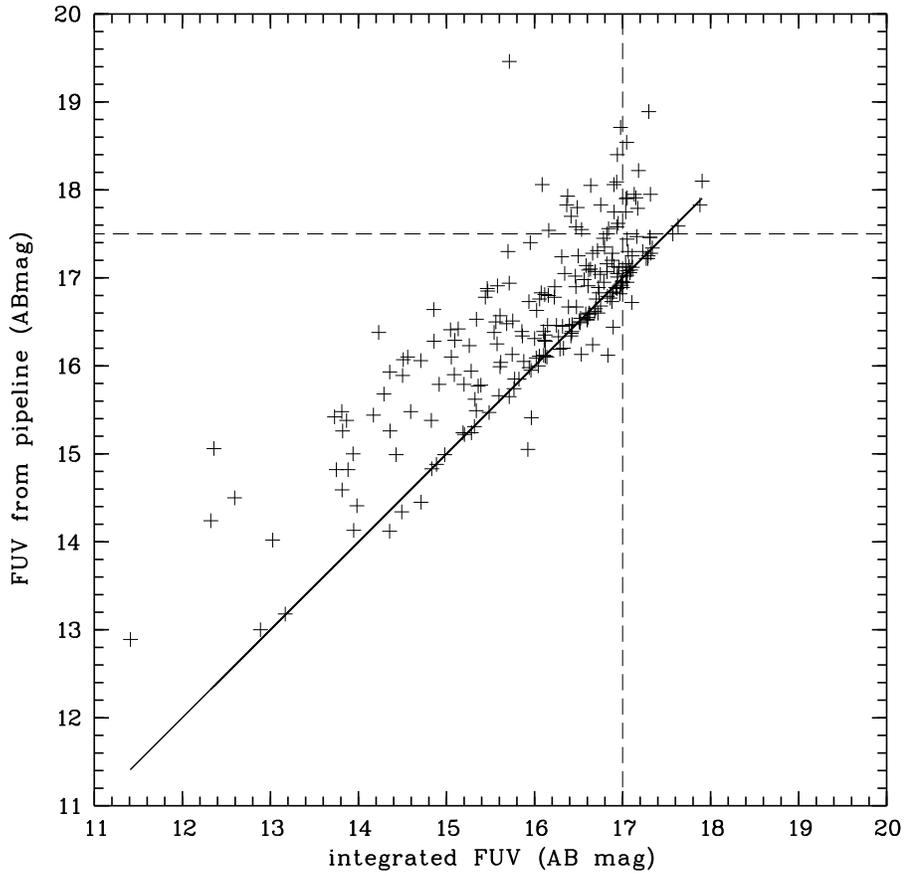}
   \caption{FUV magnitude from this paper ($x$-axis) against the FUV 
     magnitude from the pipeline (MAST archive) for the FIR-selected 
     galaxies brighter than $\mbox{FUV} = 18 $ mag. 
     The dotted lines represent  the limits applied for  the FUV selection (see text for details)}
      \label{figphot}
   \end{figure}

The selection of the FUV sources was carried out over the same area of
sky as for the FIR-selected sample (that is, excluding areas
contaminated by cirrus.)  We have checked that it is equivalent to
select lines of sight with $E(B-V)<0.08$ mag \citep[][]{schlegel}.
Our aim was to build a galaxy sample as complete as possible down to
$\mbox{FUV} = 17\;\mbox{mag}$.  Galaxies are often resolved in the FUV
and, as such, can be shredded into multiple fragments by the standard
{\sl GALEX} pipeline \citep[e.g.,][]{buat05,seibert} which is
optimized to find and extract point sources.  As a consequence we
decided to pre-select all the FUV sources brighter than
$\mbox{FUV(pipeline)} = 17.5\; \mbox{mag}$ (as estimated by the {\sl
GALEX} pipeline), where the FUV magnitudes were corrected for the
Galactic extinction before selecting at $\mbox{FUV} = 17.5
\;\mbox{mag}$.  The star/galaxy separation was made by
cross-correlating the sample with the HyperLeda and 2MASS databases.
As for the FIR-selected sample, we performed of all the FUV photometry
manually.  We estimated the level of completeness due to the
shredding.  Indeed only sources brighter than $\mbox{FUV(pipeline)} =
17.5\; \mbox{mag}$ were pre-selected because certain galaxies might, in their totality,  be
brighter than this limit but they were cataloged with a lower flux because
fainter sub-parts were detected, extracted and measured individually by the
pipeline.  We used the FIR-selected sample to quantify this effect.
We selected 238 galaxies in this sample brighter than
$\mbox{FUV(total)} = 18 \;\mbox{mag}$ where ''total" means the fully integrated
magnitude.  For these galaxies we compared the integrated FUV
magnitude that we measured manually with the FUV magnitude given by the
pipeline as posted in the MAST archives.  The result of the comparison is
plotted in Fig.~\ref{figphot}.  As expected the pipeline
under-estimates the FUV flux, sometimes with a very large factor.  Our
initial selection of the FUV sources at $\mbox{FUV(pipeline)} = 17.5
\;\mbox{mag}$ (horizontal dotted line in Fig~\ref{figphot}) ensures us
a completeness that is larger than 95$\%$ for galaxies brighter than
$\mbox{FUV(total)} = 16\; \mbox{mag}$ and of $80\;\%$ for galaxies
with $16 < \mbox{FUV} < 17\;\mbox{mag}$.  Therefore in the following
we adopt a $\mbox{FUV(total)} = 17\; \mbox{mag}$ (vertical dotted
line in Fig~\ref{figphot}) as the lower flux limit to our sample.  762 galaxies
have FUV$<$~17 mag.

The FIR fluxes of these galaxies were mainly taken from the {\sl IRAS} Faint Source Catalog 
(FSC, \citep[][]{moshir}); and if an FSC flux was  not available, it was measured by us using the Scan Processing 
and Integration Facility (SCANPI). 
This sample  was cleaned in the same way as for the FIR-selected sample: 
we accounted for confusion effects within the {\sl IRAS} beam by again searching 
for neighbours in the 2MASS and NVSS catalogs. 
Dubious cases (several candidates) were checked individually. 
18 galaxies were not detected by {\sl IRAS} at all, 123 were not detected at 60 
$\mu$m and an upper limit of $0.2\;\mbox{Jy}$ is adopted (as given in the 
{\sl IRAS} Faint Source Catalog),  39 objects  are affected by confusion. 
In the end we compiled a sample of 705 galaxies without confusion and for 
which a detection or an upper limit at 60 $\mu$m is available. 
Radial velocities and morphological types were added from HyperLeda and NED. 
All sources  have a radial velocity measurement; 51 galaxies have no 
morphological type. 
As for the FIR-selected sample we only retain the 656 galaxies with 
$v>1000 $km s$^{-1}$.  Once again, 38 galaxies are mentioned to have an active nucleus 
and 44 are classified E--S0. 
606 of the 656 galaxies have been observed by {\sl IRAS} without any confusion, 
533 were detected at $60\;\mu$m. 
$89\;\%$ of the galaxies have an $H$-magnitude in 2MASS. 
The sample of the  606 FUV-selected galaxies, without confusion and for which 
a detection or an upper limit at 60 $\mu$m is available, is presented in Table 2. 
As for the FIR-selected sample, when the galaxies are not detected at 100 $\mu$m 
we estimate this flux by using the mean value $F_{60}/F_{100}=0.4$ ($\sigma=0.2$) 
found for the galaxies in our sample detected at both wavelengths. 

\section{The luminosity and redshift distributions of the sample galaxies}

The FIR and FUV-selected samples are likely to be biased toward FIR- and 
FUV-strong emitters. 
As a consequence we expect the distribution of the FIR to FUV luminosities to 
be different within each sample, as was found in previous studies 
\citep[][]{buat05,martin,igle06}. 

\subsection{ Flux distributions}

\begin{figure}
\centering
 \includegraphics [angle=-90,width=18cm]{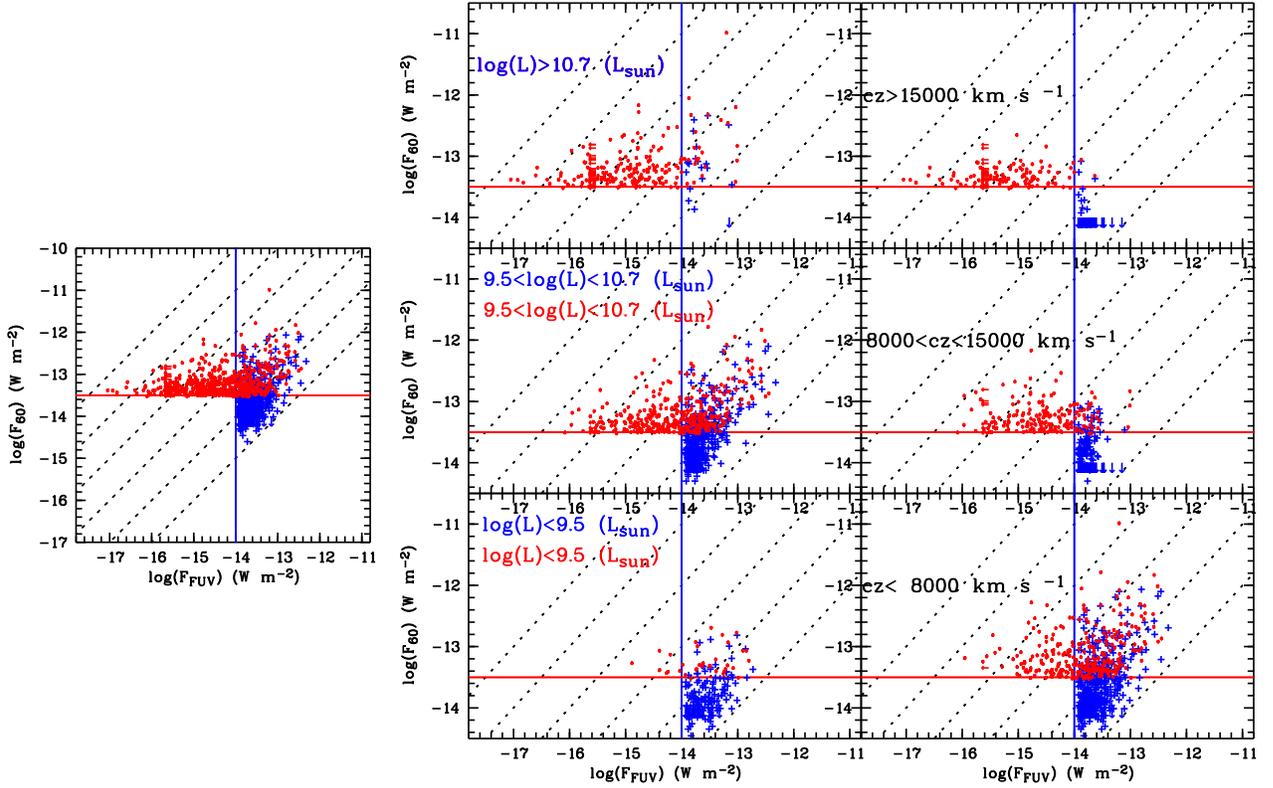}
   \caption{FUV flux against the FIR one for the FIR (red circles) and 
     FUV (blue crosses) selected samples, dotted lines: constant FIR to FUV 
     flux ratio. 
     Solid lines: FIR (horizontal line) and FUV (vertical line) selections. 
     In the right panels  the samples are splitted according to the 
     luminosity ($L_{60}+L_{\rm FUV}$) or distance (as traced by their 
     velocity) of the galaxies.  Arrows indicate upper limits. No upper limit is plotted in the selection per bin of luminosity since  $L_{60}+L_{\rm FUV}$ is defined only for galaxies detected at both wavelengths}
      \label{figfluxflux}
   \end{figure}

{}To highlight selection effects, in Fig.~\ref{figfluxflux} we 
plot $F_{60}$ versus $F_{\rm FUV}$ for the whole sample splitting them 
according to ($L_{60}+L_{\rm FUV}$) and radial velocity 
(i.e., their distances). 
The dotted diagonal lines are the loci of constant $F_{60}/F_{\rm FUV}$ 
and can thus be considered as lines of relatively constant dust attenuation 
[the derivation of a quantitative attenuation is in fact based on an 
analysis of the total infrared emission (TIR) \citep[e.g.,][]{buat05}
and not solely on  60 $\mu$m emission, but for the purpose of the present 
qualitative discussion we can safely neglect this difference]. 
{}From  Fig.~\ref{figfluxflux}, it is obvious that the FIR selection 
focuses on extinguished galaxies whereas the FUV selection is biased toward 
galaxies with a low $F_{60}/F_{\rm FUV}$ ratio.  
Indeed the FIR-selected sample  exhibits a long tail towards large 
$F_{60}/F_{\rm FUV}$ ratios. 
More distant galaxies are selected in the FIR than in the FUV. 
\citep[e.g.,][and discussion below]{igle06,xu06}.

\subsection{ Mono-variate luminosity functions}

\begin{figure}
\centering
\includegraphics[width=9cm]{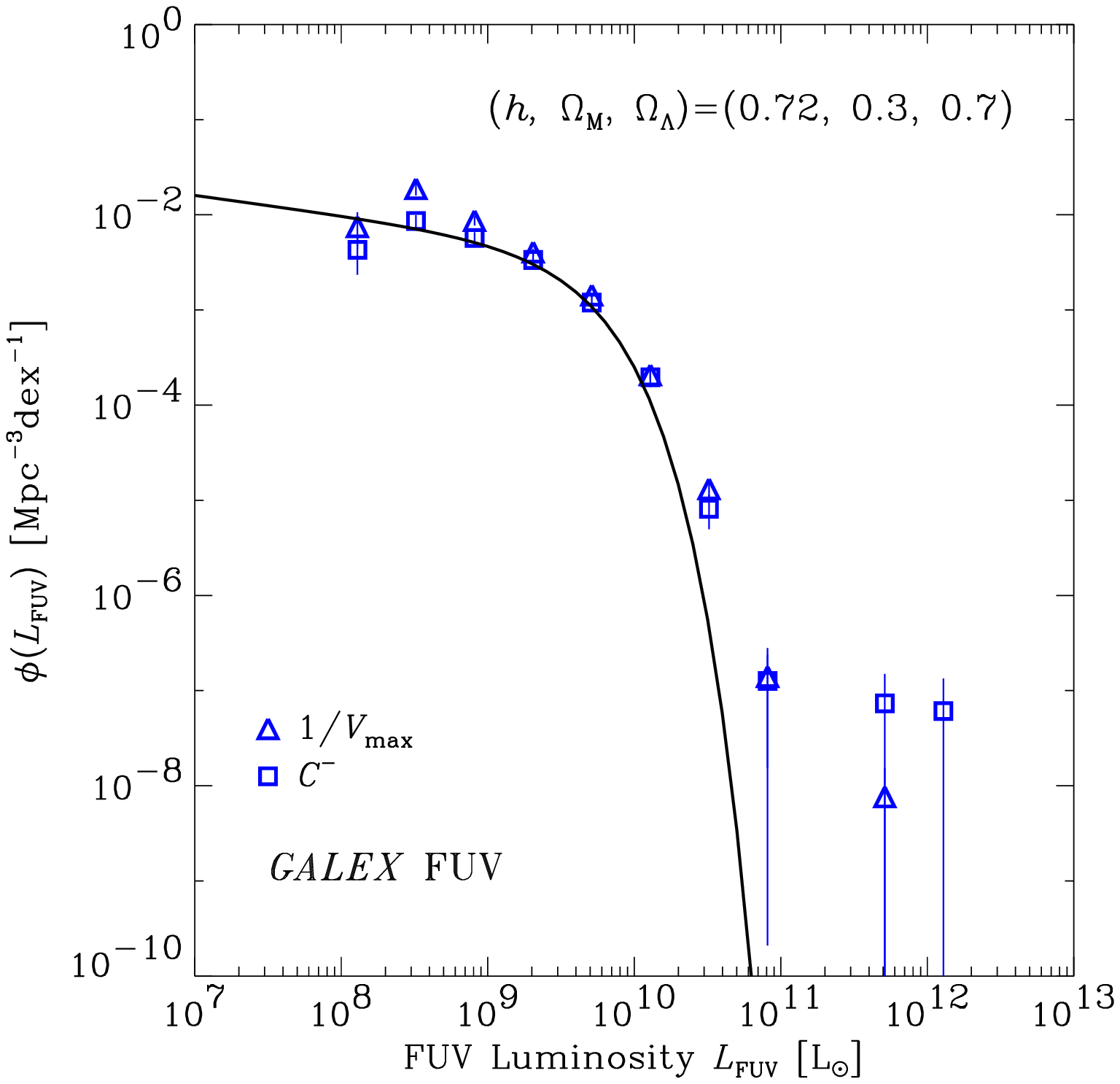}
\includegraphics[width=9cm]{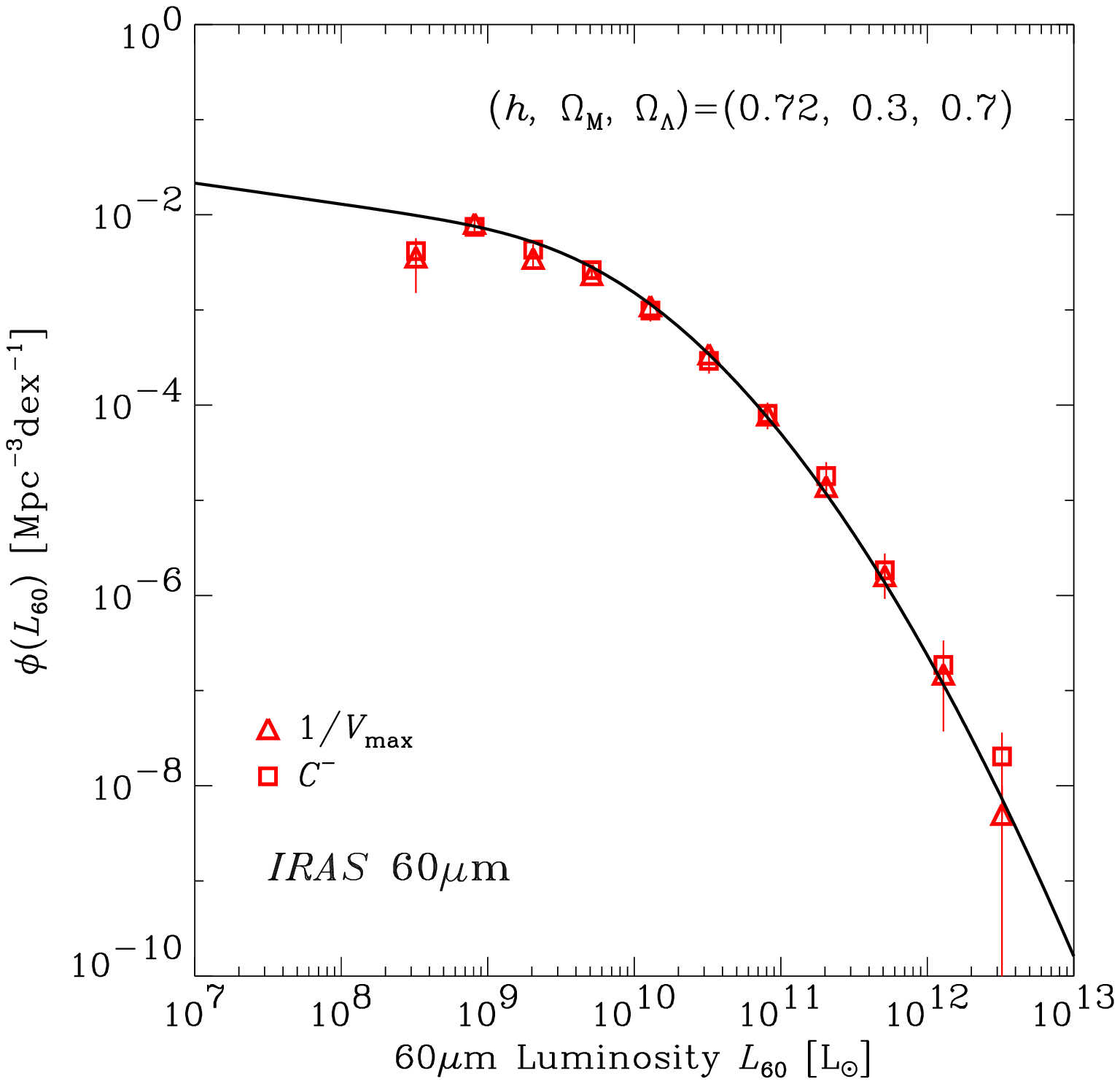}
   \caption{FUV and 60 $\mu$m Luminosity Functions for the FUV and FIR 
     selected samples. 
     The  LF are estimated by $1/V_{\rm max}$-method 
     (open triangles) and $C^-$-method (open squares).
     We also overplot the analytic FUV LF from \citet[][]{wyder} 
     and  the 60 $\mu$m LF of \citet[][]{ttt} (solid lines).
     Errors (1 $\sigma$) are calculated by bootstrap resampling.
     }
      \label{monoLF}
   \end{figure}

The very different distributions found in Fig.~\ref{figfluxflux}  are 
the direct consequence of the shapes of the individual FIR and FUV luminosity functions 
(LF). These are known to be very different \citep[e.g.,][]{buat98,ttt2,igle06}. 
We have calculated the FUV and 60 $\mu$m luminosity functions for our FUV 
and FIR-selected samples respectively using the $1/V_{\rm max}$ and the 
$C^-$ methods, using the recipes of \citet{takeuchi00a}. 
We excluded galaxies known to be active since we are only interested in a  measurement 
of the star formation (non-AGN) activity from the FIR and the FUV emissions. 
The optimal bin size is determined systematically by the formula of 
\citet{takeuchi00b}, and the error bars are obtained by boot-strap resampling
\citep[for details, see ][]{takeuchi00a}.
We checked that $K$-corrections are negligible; the study is 
restricted to the very  nearby Universe \citep[cf.][]{wyder}. We examined the completeness of the sample also in the LF analysis,
using the number counts of galaxies.
Our sample again turned out to be highly complete, brighter than
17~mag for the UV and brighter than 0.6~Jy for the FIR sample.
The LFs are reproduced in Fig.~\ref{monoLF} and compared to the local 
LFs built by \citet[][]{wyder} in FUV and \citet[][]{ttt} at 60$\mu$m. 
The agreement is very good, thereby encouraging us to believe that our samples are  
representative of nearby-universe populations. For both selections, the faintest 
bins appear to be under-estimated. In the following we will exclude these bins from the analysis.
The very similar results found with both methods ($1/V_{\rm max}$ and the 
$C^-$) validates the use of $V_{\rm max}$ in the following analyses.

The extension of the 60$\mu$m LF toward bright galaxies, as compared to 
the FUV one, implies that we see intrinsically brighter and more distant 
galaxies in  an FIR-selected sample.

\section{Do we see the same galaxies in FIR and in FUV?}

One of the fundamental question to address is: do we see the same universe 
in the FIR and FUV, or do we miss galaxy populations when working at a single 
wavelength? 
For example, \citet{adelberger} claimed to see all the star formation 
at $z=3$ from a purely UV selection criterion.
At low redshift,  \citet[][]{xu06} concluded that we globally see the same 
galaxies but the statistics were poor, and subtle effects might have been 
difficult to examine.
We will now re-examine these questions with our much larger samples.
Given the very different shapes of the luminosity functions and the different volumes 
sampled by each selection (FIR and FUV) the comparison is not trivial 
\citep[e.g.,][]{xu06}. 
We must also define intrinsic properties independent of the wavelength 
selections to compare the samples. 
Since we are interested in the measure of the star formation we will 
focus on the distribution of bolometric luminosities from young stars.

\subsection{ Bolometric luminosity functions from young stars} 

A very crude and spread-out way of estimating the total energy coming from 
newly formed stars is simply to add $L_{\rm FUV}$ and $L_{60}$ 
\citep[e.g.,][]{martin,wang,xu06}. 
We can also make a more sophisticated analysis by first calculating the total 
IR luminosity ($L_{\rm TIR}$) to account for all of the emission from dust.  
In addition to the better physical significance of $L_{\rm TIR}$ as compared to 
$L_{60}$, the post-{\sl IRAS} infrared observations ({\sl ISO}, {\sl SPITZER}, 
{\sl AKARI}) are made at different wavelengths from 
{\sl IRAS}, and there again the comparison is usually made through the TIR emission. 
Therefore $L_{\rm TIR}$  is better suited than $L_{60}$ for a comparison with 
studies at higher redshift. 
We calculate $L_{\rm TIR}$ by combining the fluxes at 60 and 100 $\mu$m 
according to the recipes in \citet{dale}. 
$L_{\rm FUV}$ calculated as $\nu L_{\nu}$ is a good measure of the
total luminosity between 1200 and 3600~\AA\ as long as no dust
attenuation occurs: using Starburst99 under the hypothesis of a
constant star formation rate over 100 Myr leads to
$L(\mbox{1200--3600~\AA}) \sim 0.8 L_{\rm FUV}$.  Nevertheless, a dust
attenuation which is strongly dependent on the wavelength may induce
large variations in this relation: as an example using
\citet[][]{calzetti}'s attenuation law and a color excess $E(B-V) =
0.3$ (corresponding to $A(\mbox{FUV})= 3\; \mbox{mag}$ and a TIR to
FUV flux ratio of $\sim 30$) implies $L(\mbox{1200--3600~\AA}) \sim
1.3 L_{\rm FUV}$.  \citet[][]{hirashita} showed that the effect on the
total estimate of the star formation rate is not large as long as the
contribution of the TIR and FUV emissions are added (the uncertainties
are reduced by the large contribution of the TIR emission to the total
bolometric one).  Following \citet[][]{hirashita}, we also account for
the dust heating by stars older than ~100 Myr which generates a TIR
emission not related to the recent star formation: the total
luminosity of young stars is thus expressed as $L_{\rm bol} = L_{\rm
FUV} +(1-\eta) L_{\rm TIR}$ where $\eta$ is the fraction of the TIR
emission not related to the star formation.  The value of $\eta$ is
found to be between 0.2 and 0.4 in the nearby universe
\citep[][]{bell03,hirashita,igle06}.  Hereafter we adopt $\eta =0.3$.
The total luminosity of young stars can therefore be written as
$L_{\rm bol} = L_{\rm FUV} +0.7 L_{\rm TIR}$.  This formula is
established for local star forming galaxies and confirmed to be valid
for nearby galaxies selected in the FIR or FUV
\citep[][]{igle04,igle06}. Nevertheless, in galaxies forming stars
very actively, the entire TIR luminosity is likely to be related to
the recent star formation in starburst galaxies
\citep[e.g.,][]{hirashita}).

The bolometric luminosity function from young stars is calculated within 
each sample using the $1/V_{\rm max}$ weighting method. 
We test the influence of  $K$-corrections by interpolating the FUV and 
NUV magnitudes according to the redshift, they are found negligible in 
both samples. 
As for the derivation of the  mono-variate luminosity functions (section 3.2), galaxies known to be active are dropped. 
We consider the entire sample including upper limits. 
Therefore the $V_{\rm max}$ value of each galaxy is defined by its 
luminosity and the detection limit relevant for the  selection (FUV or FIR)
since the volume sampled by the other wavelength is infinite as soon as 
upper limits are included. 
The adopted limits are that of the {\sl IRAS} Point Source Catalog for 
the FIR-selected sample (0.6~Jy at $60\;\mu$m) and $\mbox{FUV} = 17 
\;\mbox{mag}$ for the FUV-selected sample (by construction).
{}Taking into account the upper limits, we consider 
two extreme scenarios in  calculating the luminosity functions:
Scenario 1: upper limits are considered as true detections, 
Scenario 2: a flux equal to zero is adopted for a non-detection.
 
\begin{figure}
\centering
 \includegraphics [angle=-90,width=15cm]{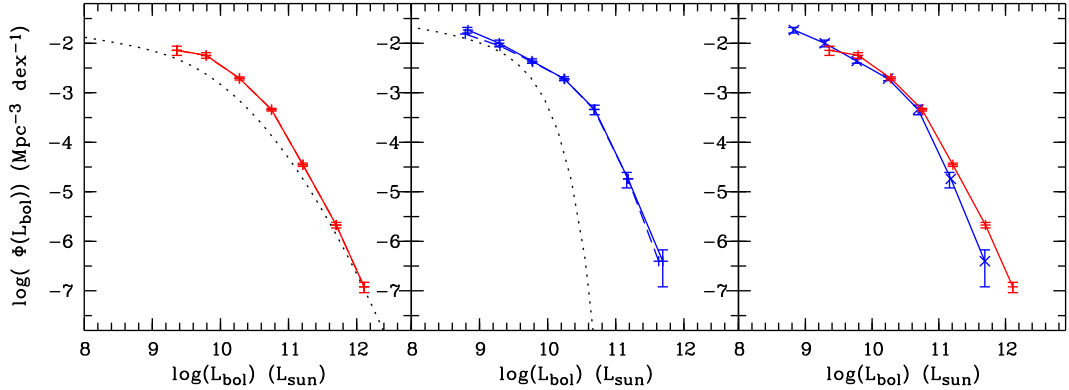}
\caption{Bolometric luminosity functions calculated for the two considered scenarii (scenario 1: solid line,  
  scenario 2: dashed line), $L_{\rm bol}$ is defined as $L_{\rm bol} = L_{\rm FUV} +(1-\eta) L_{\rm TIR}$, the 1 $\sigma$ error bars are overplotted. Left panel:  FIR-selected sample, the monochromatic (60 $\mu$m) LF from \citet[][]{ttt3} is represented 
  as  a dotted line; middle panel: FUV-selected sample, the  monochromatic (FUV) LF \citep[][]{wyder} is represented as 
  a dotted line; right panel: comparison of the bolometric LF for the FUV-selected and FIR-selected 
  samples: crosses (X) and blue line are used for the FUV selection and plus symbols (+) and red line for the FIR selection. Only the scenario 1 is considered.}
  
\label{UVselIRselLFtot}
\end{figure}

In Fig~\ref{UVselIRselLFtot}  we plot the bolometric 
LF $\phi(L_{\rm bol})$ for each sample as compared to the monochromatic ones 
at 60 $\mu$m \citep{ttt3}  and at 1530~\AA \citep{wyder}.

The two scenarios adopted to calculate the bolometric LF agree very  
well for the FIR-selected sample.  
It therefore appears that the 60 $\mu$m luminosity is a robust tracer of the 
luminosity of young stars: the bolometric LF appears to be  shifted 
as compared to the  LF at 60 $\mu$m by at most a factor of  $\sim 3$ (for 
intermediate luminosities). The difference between LFs decreases as the luminosity increases: 
for the highest luminosities the 60 $\mu$m luminosity function is similar to the
bolometric one.

The two scenarios adopted to calculate the bolometric LF in the FUV-selected 
sample also agree quite well: slight differences (lower than 0.2 dex in the vertical 
direction and 0.1 dex in the horizontal) are visible at the faint and luminous ends; 
but they are smaller than the one-$\sigma$ errors calculated for each scenario.

Even within a FUV selection the FUV flux alone (without any correction) 
misses a large part of the total emission of FUV-selected galaxies. Whereas the FUV 
flux appears to be a reliable estimator of the bolometric emission from young stars in 
low luminosity galaxies ($L_{bol} < 2.5~ 10^9 L_\sun$), the difference increases very 
fast with luminosity: the FUV luminosity is  $\sim$ 5 times lower than the bolometric 
one for $L_{bol} \sim 10^{10} L_\sun$ and  the discrepancy reaches a factor $\sim 500$ 
for $L_{bol}= 3 ~ 10^{10} L_\sun$. This trend has to be related to the relation found 
between the luminosity (or star formation rate) of galaxies and their dust attenuation 
\citep[][and discussion in section 4.3]{wang,buat98,hopkins01,sullivan}.
As a consequence, large luminosity-dependent corrections must be applied to 
the FUV emission in order to retrieve all of the bolometric luminosity of  young stars 
(i.e., the recent star formation rate) of FUV-selected galaxies. 

In the following, and in order to simplify the discussions, we will adopt 
Scenario 1 for the calculations; that is, we will include upper limits as 
detections. 

Fig ~\ref{UVselIRselLFtot} compares the bolometric LF for both
samples. It can be seen that both luminosity functions are consistent
for intermediate luminosities: in the nearby universe these galaxies
are detected equally well in FIR and in FUV.  For bolometric
luminosities larger than $5 ~ 10^{10} L_\sun$ the bolometric LF
derived from the FIR selection is higher than that derived from the
FUV, and the discrepancy increases with the luminosity: we miss
intrinsically bright galaxies which appear much fainter in FUV
(cf. section 4.3). The shallowness of the {\sl IRAS} survey does not
allow us to compare the distributions at low luminosities ($L_{bol} <
\sim 2~ 10^9 L_\sun$).

\subsection{Energy distributions}

\begin{figure}
\centering
 \includegraphics [angle=-90,width=18cm]{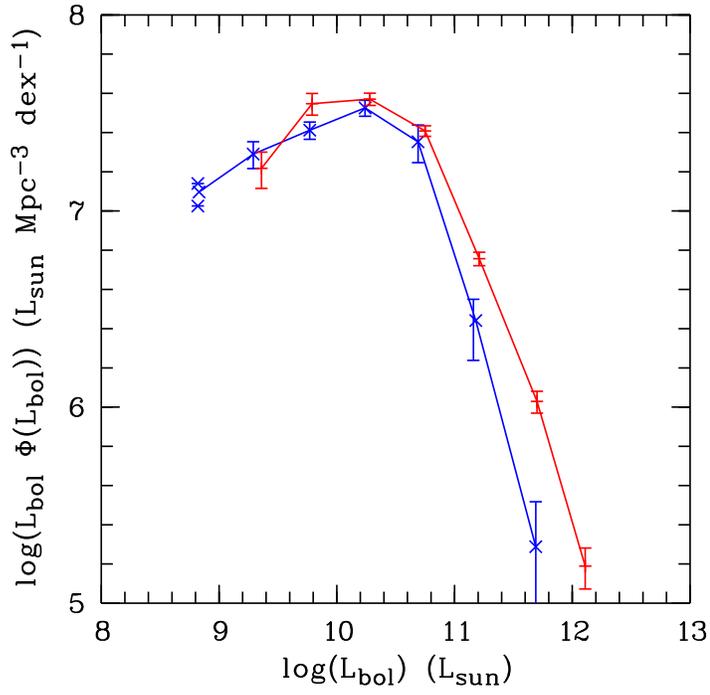}
\caption{Energy distribution ($\log(L_{\rm bol}\phi(L_{\rm bol}))$) as a function of $\log(L_{\rm bol})$ for the FUV and the FIR 
  selected  samples. Same symbols  as in Fig~\ref{UVselIRselLFtot}. The 1$\sigma$ error bars are also overplotted.
}
\label{LphiL}
\end{figure}

In order to  estimate what fraction of the  energy emitted by young stars in the nearby universe is recovered from 
an FUV selection and how much from an FIR selection in, we have calculated 
the  $L_{\rm bol} \phi(L_{\rm bol})$ product which represents the energy 
contribution of galaxies with a given  $L_{\rm bol}$ luminosity to the 
luminosity density in the local universe as a whole. 
The distributions are shown in Fig~\ref{LphiL} 
(all the calculations are made with  Scenario 1, i.e. with  upper 
limits considered as detections). 
The distributions are  marginally consistent at the one-$\sigma$ level 
at intermediate luminosity. 
As for the luminosity functions (section 4.1) a discrepancy  appears at high 
luminosity and increases with the luminosity:  at  $L_{\rm bol}\sim 5~10^{10} L_\sun$ the FUV selection 
systematically under-estimates the luminosity density by a factor $\sim 1.5$ 
 and the factor  reaches $\sim 5$ for $L_{\rm bol} \sim 3 ~ 10^{11} L_\sun$. \\
At the faint end it seems that the situation is reversed, with a large number of low-luminosity 
UV-selected galaxies not being present in our FIR selection. Deeper FIR 
observations will be necessary to  confirm or negate this trend:  future ASTRO-F/AKARI 
observations will allow us to address this issue.

\subsection{$L_{\rm TIR}/L_{\rm FUV}$ distributions}

Analysing the $L_{\rm TIR}/L_{\rm FUV}$ ratio is  another way to 
compare the FIR and FUV selection effects \citep[][]{martin,xu06}.
Indeed, the  $L_{\rm TIR}/L_{\rm FUV}$ ratio has a physical significance 
since it is directly related to the dust attenuation in star-forming 
galaxies \citep[e.g.,][]{buatxu,gordon,buat05}. 
This ratio gives us information about the dust obscuration as well as 
the differences and/or similarities in  the galaxies selected in different 
ways (FIR versus FUV). 
Hereafter we will deal with the $L_{\rm TIR}/L_{\rm FUV}$ ratio taking 
in mind that it can be calibrated in absolute dust attenuation at FUV 
wavelengths using for example the formulae of \citet[][]{buat05}:
\begin{eqnarray}
 A\mbox{(FUV) [mag]} &=& -0.0333~\left(\log \frac{L_{\rm TIR}}{L_{\rm FUV}}
   \right)^3 \nonumber \\
   &&+0.3522~\left(\log\frac{L_{\rm TIR}}{L_{\rm FUV}}\right)^2\nonumber \\
   &&+1.1960~\left(\log\frac{L_{\rm TIR}}{L_{\rm FUV}}\right)+0.4967 \;.
\end{eqnarray}

\begin{figure}
\centering
 \includegraphics [angle=-90,width=18cm]{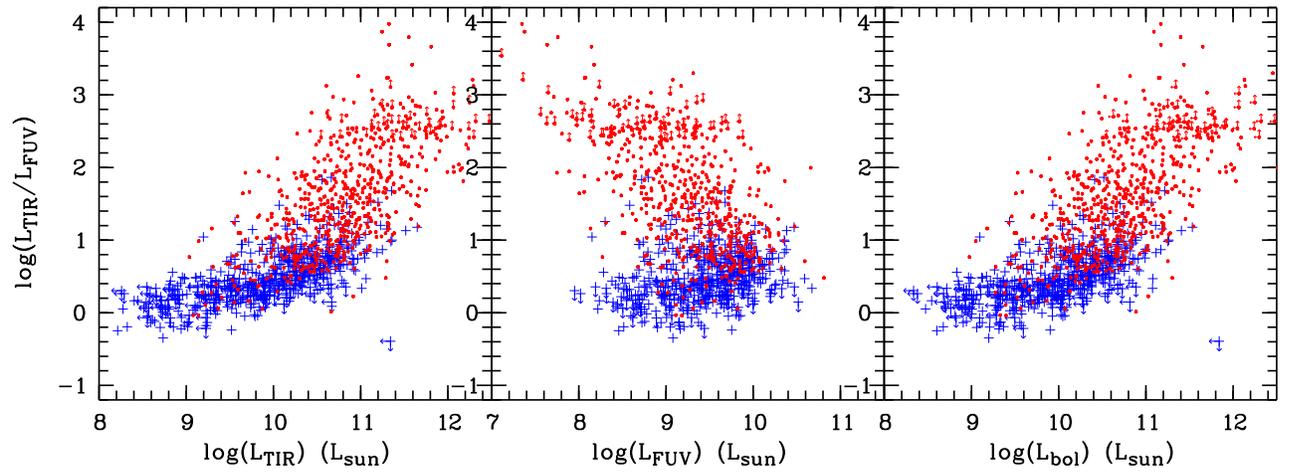}
   \caption{$L_{\rm TIR}/L_{\rm FUV}$ ratio versus 
     $L_{\rm TIR}$, $L_{\rm FUV}$ and $L_{\rm bol}$ for the FIR-selected 
     sample (circles) and the FUV-selected sample (crosses). 
}
      \label{figltir_lfuvplot}
   \end{figure}

\begin{figure}
\centering
 \includegraphics [angle=-90,width=18cm]{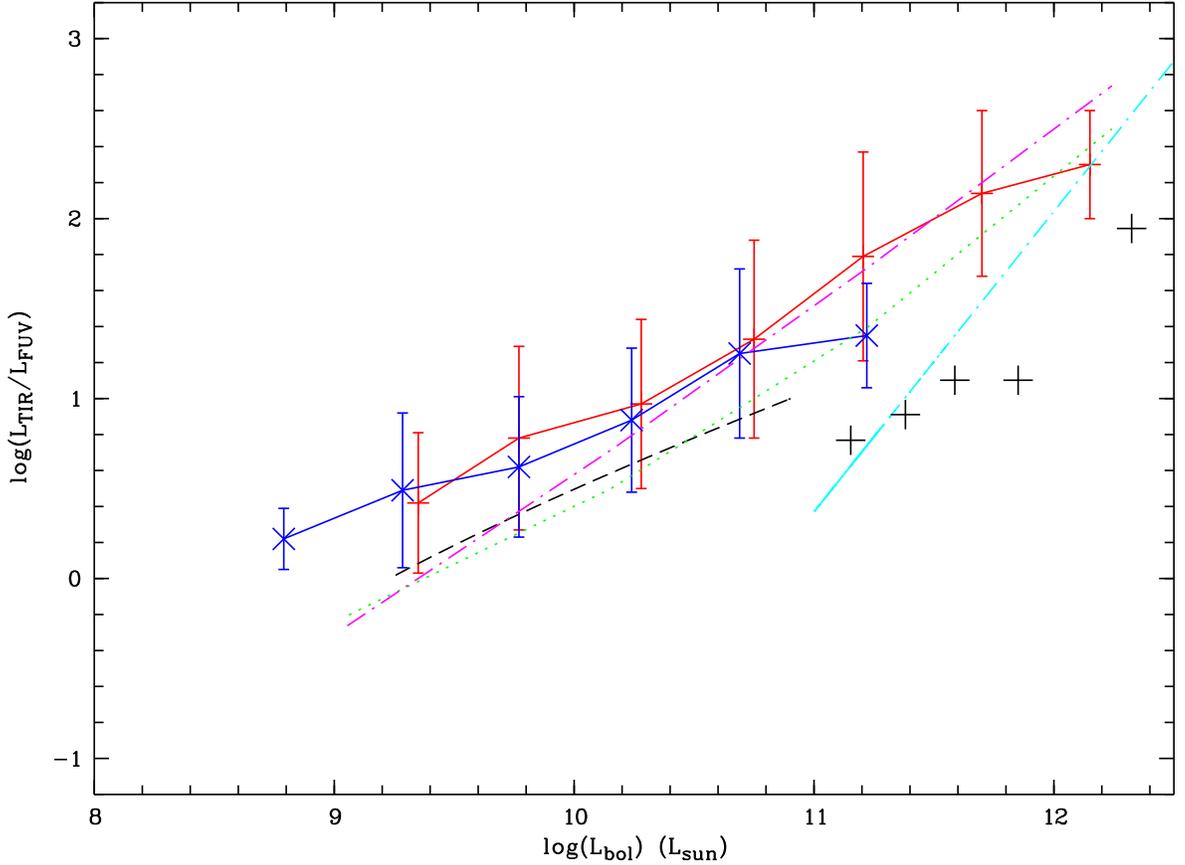}
   \caption{Mean $L_{\rm TIR}/L_{\rm FUV}$ ratio versus $L_{\rm bol}$ 
     calculated for the FUV-selected sample (blue solid line and crosses)  and  for the 
     FIR-selected sample (red solid  line and plus (+) symbols), the errors  (1$\sigma$) are 
     overplotted as vertical bars. 
     The green dotted line is from \citet[][]{martin}, 
     the dot-dashed magenta line from \citet[][]{xu06}, 
     the black dashed line from \citet[][]{bell03}, 
     the dot-dashed cyan line from \citet[][]{reddy} (optically selected galaxies at z$\sim$2 also observed at 24 $\mu$m)  and the crosses correspond to mean values per bin of luminosity  for LBGs at z$\sim$1 from \citet{burgarella2}
}
      \label{figltir_lfuvfit}
   \end{figure}

Fig.\ref{figltir_lfuvplot} shows the variation of $L_{\rm TIR}/L_{\rm
FUV}$ as a function of $L_{\rm bol}$ for the two samples under
consideration (FIR and FUV-selected).  It is also useful to consider
the variation of this ratio as a function of the ''monochromatic"
luminosities (at 60 $\mu$m, or in the FUV band alone). These plots are
found in Fig. \ref{figltir_lfuvplot}.  $L_{\rm TIR}/L_{\rm FUV}$
(i.e., the dust attenuation) is found to increase with $L_{\rm TIR}$
and with $L_{\rm bol}$ in both samples.  Such an increase of $L_{\rm
TIR}/L_{\rm FUV}$ with the TIR luminosity confirms previous results
\citep[]{wang,buat98,hopkins01,sullivan} and appears to be robust
against selection effects.  The similarity between the trends found
with $L_{\rm TIR}$ and $L_{\rm bol}$ is obvious and is due to the
dominant contribution of the TIR luminosity to the bolometric
luminosity, as compared to the FUV contribution
(cf. Fig.~\ref{UVselIRselLFtot}).  The trend is steeper and more
scattered for the FIR selection than for the FUV.  Very different
trends are found within each sample with $L_{\rm FUV}$: a strong
decrease of $L_{\rm TIR}/L_{\rm FUV}$ with $L_{\rm FUV}$ is observed
for the FIR-selected sample whereas a very loose positive correlation
is found for the FUV selection (correlation coefficient equal to 0.2)
making irrelevant any correction of the dust attenuation based on the
observed FUV luminosity alone.  It confirms the results of
\citet[][]{igle06} that the FUV emission emerging from galaxies
selected in the FIR represents only a very small fraction of the total
luminosity emitted by the young stars.  It is worth noting that
galaxies selected with a very high FUV luminosity ($ \ga 2 ~ 10^{10}
L_\sun$) exhibit a rather moderate $L_{\rm TIR}/L_{\rm FUV}$ ratio
(i.e. a small attenuation: $A(FUV) = 1.5\pm 0.5$ mag using eq. (1)),
some of them having a dust attenuation as low as $\sim$0.5 mag
corresponding to $L_{\rm TIR}/L_{\rm FUV}\sim 1$. These galaxies are
called UV luminous galaxies (UVLGs) by \citet{heckman}. A subclass of
these galaxies is thought to contain the  analogs of the more distant Lyman
Break Galaxies \citep[][and discussion below]{hoopes}. As noted in the
previous section, we must also account for the different volumes
explored within each selection in order to secure the trends.  To this end we
now calculate weighted distributions of $L_{\rm TIR}/L_{\rm FUV}$ as
a function of $L_{\rm bol}$.  We divide the sample into bins of $L_{\rm
bol}$ ($\mbox{bin size} = 0.5\; \mbox{dex}$) and for each bin we
calculate the averaged ratio $R = \log (L_{\rm TIR}/L_{\rm FUV})$ and its
standard deviation as follows:
\begin{equation}
  \langle R(L_{\rm bol}) \rangle =
  \frac{\sum_i \omega_i R_i}{\sum_i{\omega_i}}
\end{equation}
and 
\begin{equation}
  \sigma^2(L_{\rm bol})= \frac{\sum_i\omega_i (R_i- \langle R(L_{\rm bol})
    \rangle)^2}{\sum_i{\omega_i}}
\end{equation}
where $\omega_i$ is the weight for the $i$-th galaxy, practically 
$1/V_{\rm max}$. 
$V_{\rm max}$ is calculated as for the bolometric luminosity 
functions under the Scenario 1 (with upper limits treated as true detections). 
Indeed,  Scenario 2, which essentially defines the non-detected sources as having zero flux, is irrelevant to the analysis of the 
$L_{\rm TIR}/L_{\rm FUV}$ ratio. 
Adopting Scenario 1 is a conservative approach when searching for 
differences between the FIR and FUV-selected samples: the non-detections 
lead to upper limits for $L_{\rm TIR}/L_{\rm FUV}$ in the FUV-selected 
sample and lower limits in the FIR-selected sample. 
The results of the calculations are plotted in Fig~\ref{figltir_lfuvfit}.
Both samples give similar trends at low and intermediate luminosities, 
but the volume corrections cannot completely compensate  for the very different 
distributions seen in Fig.~\ref{figltir_lfuvplot} for the high luminosities. 
Whereas the $L_{\rm TIR}/L_{\rm FUV}$ ratio continues to increase with 
luminosity for FIR-selected galaxies, it shows a clear flattening for 
FUV-selected galaxies brighter than $5 ~ 10^{10}~L_\sun$, 
and the $L_{\rm TIR}/L_{\rm FUV}$ seems to reach an asymptotic value 
which corresponds to a dust attenuation $A(\mbox{FUV}) \simeq 2.5$ mag.

For the nearby universe, \citet[][]{bell03} analyzed a  sample of nearby 
galaxies  and found that $L_{\rm TIR}/L_{\rm FUV} \simeq 
(L_{\rm TIR}/10^9)^{0.5}$, where $L_{\rm TIR}$ is expressed in solar units. 
The galaxies selected by Bell have $L_{\rm TIR} <10^{11} L_\sun$. 
His mean relation (transformed to the quantities used here: 
$L_{\rm TIR}/L_{\rm FUV}$ and $L_{\rm bol}$) is shown in Fig.\ref{figltir_lfuvfit}. The 
general behavior is similar to that found for our FUV selection, where
the Bell relation gives lower $L_{\rm TIR}/L_{\rm FUV}$ for a given 
luminosity but still within our  one-$\sigma$ error bars. 
We can also compare our results to those also obtained from 
a {\sl GALEX}/{\sl IRAS} comparison \citep[][]{martin,xu06}. 
{}To perform this comparison we have transformed the 60 $\mu$m luminosity 
used in those works into a TIR luminosity by applying a factor 2.5 
\citep[][]{ttt3,takeuchi06}. 
\citet[][]{xu06} did not find a significant difference between the FIR 
and the FUV selection. The mean relation they found is overplotted 
in Fig \ref{figltir_lfuvfit}. 
This is consistent with the present analysis, especially 
for intermediate luminosities. 
The consistency is only marginal at low luminosity 
($L_{\rm bol} \la 5 \times 10^{9} L_\sun$) and for the last bin  
in our FUV-selected sample with $L_{\rm bol} >10^{11} L_\sun$: we obtain 
larger values of $L_{\rm TIR}/L_{\rm FUV}$ than Xu et al.
These differences can be explained if we go back to the sample selections. 
The samples used by Xu et al. were  smaller and shallower: 
the FUV-selected sample contained only 94 objects brighter than  $\mbox{NUV}
= 16\; \mbox{mag}$ and the FIR-selected sample had only 161 galaxies 
(with a similar selection as in the present study). 
As a consequence, the high- and low-luminosity ranges are likely to be 
undersampled in  the Xu et al. study. Another difference comes from the treatment 
of confused objects (i.e., galaxies not resolved by {\sl IRAS}, cf. section 2): Xu 
et al. included them in their analysis whereas they are excluded from the 
present work. A reliable study of the difference of behavior between isolated 
and close pairs/interacting galaxies will be addressed with future ASTRO-F/AKARI 
data whose spatial resolution will be much better than that of {\sl IRAS}. 
\citet[][]{martin} used a combined sample (galaxies selected in FUV or 
at 60 $\mu$m) and analysed the $L_{\rm 60}/L_{\rm FUV}$ distribution. 
Their mean relation (translated in $L_{\rm TIR}/L_{\rm FUV}$ and $L_{\rm bol}$ 
according to the definitions we adopt in this paper) is also overplotted in 
Fig~\ref{figltir_lfuvfit}. 
If we take into account the dispersions found in  both  studies (only ours are 
reported on the figure, but those of Martin et al.\ are similar) 
the results are consistent, although our mean values for 
$L_{\rm TIR}/L_{\rm FUV}$ are systematically higher than those obtained by 
Martin et al. 
At low luminosities, the relations begin to diverge. 
In this luminosity range the FUV selection is likely to dominate the 
Martin et al. sample and their sample  was built on an area 
3 times smaller than ours, so we can expect some undersampling of these 
bins in their study.

It is also very interesting to use our results to search for a redshift 
evolution of the dust attenuation at a given bolometric luminosity. 
The comparison is not easy because only few high redshift studies are based 
on accurate determinations of the rest-frame FUV and FIR luminosities. 
Nevertheless, with the advent of the {\sl SPITZER} data the situation is 
evolving fast. 
\citet[][]{reddy} studied optically selected $z\sim 2$ galaxies which were also observed 
by {\sl SPITZER} at 24$\mu$m.  Although the extrapolation from the observed 
MIR range to the total IR is not straightforward they compare 
the $L_{\rm TIR}/L_{\rm FUV}$ (their FUV at 1600 $\AA$ is very similar to our 
FUV band) with $L_{\rm TIR}+L_{\rm FUV}$. 
The best fit they obtain is reproduced in Fig.~\ref{figltir_lfuvfit} 
(for $L_{\rm bol}>10^{11} L_\sun$ since they  have only access to these 
galaxies). 
For $10^{11}<L_{\rm bol}< 2.5 \times 10^{11} L_\sun$ the dust 
attenuation seems to be lower at $z=2$ than at $z=0$ as claimed by 
Reddy et al. 
But for intrinsically brighter objects, the $L_{\rm TIR}/L_{\rm FUV}$ ratios
found by Reddy et al.\ are consistent with what we find in our FIR and 
FUV-selected samples without invoking any evolution of the dust attenuation. 
\citet[][]{burgarella2} selected Lyman Break Galaxies at $z\sim 1$ in 
the Chandra Deep Field South using {\sl GALEX} data. 
One-fourth of these galaxies have a {\sl SPITZER} detection at 24 $\mu$m. 
The mean values of $L_{\rm TIR}/L_{\rm FUV}$ obtained for these galaxies per bin of 
bolometric luminosity (bin size=0.5 dex) are shown in Fig.~\ref{figltir_lfuvfit}.
For these galaxies the dust attenuation (estimated through their 
$L_{\rm TIR}/L_{\rm FUV}$ ratio) is found to be consistent with that found 
by Reddy et al.\ for the same range of bolometric luminosities.
A more complete comparison of the samples is forthcoming (Burgarella et al. in preparation).\\

Both analyses of the bolometric luminosity functions and of the
$L_{\rm TIR}/L_{\rm FUV}$ ratio lead to the conclusion that a FUV
selection misses some of the most heavily obscured and intrinsically
brightest galaxies. Conversely, an FIR selection probably
underestimates the contribution of intrinsically faint (in a
bolometric sense) galaxies. Deeper FIR surveys are needed to confirm
this effect.  Our analysis is performed on galaxy samples excluding
active and confused sources. Although only few sources were excluded,
the contribution of interacting systems and close pairs must be
investigated with future AKARI data.

At higher redshift the general trend toward high-luminosity systems,
together with the increase of $L_{\rm TIR}/L_{\rm FUV}$ with the
luminosity may argue for a gradual increase with redshift of the loss
of star formation in FUV surveys. Nevertheless this effect may be
compensated for, at least in part, if the dust attenuation of UV-optical
selected galaxies substantially decreases at high z, as suggested by
\citet[][]{reddy}. Statistical studies of FIR and FUV-selected samples
have to be performed at higher z to investigate this issue. The {\sl
SWIRE/GALEX} comparison performed by \citet{xu06apjs} leads to no
apparent difference between $L_{\rm TIR}/L_{\rm FUV}$ ratios of
star-forming galaxies between z=0 and z=0.6. An analysis of deep
fields observed by {\sl GALEX} and {\sl SPITZER/MIPS} is underway.

\section{Specific star formation rates}

Since we are dealing with star formation rates, galaxies classified as 
early-type (E-S0) are excluded from the following analysis. Galaxies known to 
have an active nucleus are again excluded (as in section 3 and 4)
The present star formation efficiency of  galaxies can be quantified by  
comparing their whole stellar mass to their  present SFR. The specific SFR 
(SSFR) is defined as the ratio of the present SFR to the stellar mass:  
$ SSFR= SFR/M_\star$. 
This SSFR is closely related to the so-called $b$ parameter defined as 
the ratio of the present to past averaged SFR: 
\begin{eqnarray}
  b = \frac{\mbox{SFR}}{\langle \mbox{SFR} \rangle}
    = \frac{t\,(1-R)\,\mbox{SFR}}{M_\star}
\end{eqnarray}
where $R$ is the fraction of recycled gas, taken usually to be between
0.3 and 0.5, and t the age of the galaxies is assumed to be $\sim 13$
Gyr \citep[e.g.,][]{boselli}).  The total star formation rate is
calculated by combining the TIR and FUV derived SFR as done earlier by
\citet[][]{hirashita} and more recently by \citet[][]{igle06} for
galaxies selected in a very similar way as in this work:
\begin{eqnarray}
  \mbox{SFR}_{\rm tot} = \mbox{SFR}(\mbox{FUV}_{\rm obs})+(1-\eta)
    \mbox{SFR}(\mbox{TIR})
\end{eqnarray}
with $\eta=0.3$. 
To undertake this analysis we need to estimate the total stellar mass in our
galaxies. 
Most of them were detected by 2MASS therefore we decide to use their 
$H$-band luminosities to measure their stellar content (very similar results 
are found when using the $K$-band). 
\citet[][]{belljong} have analysed the variation of stellar mass-to-light ratios as a function of various color indices. 
We have obtained the $(B-V)$ color from HyperLeda for about 1/5 of our  galaxies
 and obtain an average value of $\langle B-V \rangle =0.6\;\mbox{mag}$ ($\sigma$=0.2 mag). 
This corresponds to  $M/L_{H}= 0.57$ (in solar units); and
we compute the stellar mass of the galaxies in our samples using this 
mean $M/L_{H}$. The uncertainty is estimated to be $\sim 30 \%$ if we account 
for a standard dispersion on (B-V) of 0.2 mag (leading to $M/L_{H}$ 
falling between 0.4 and 0.8). 
The completeness in $H$ is very high in our samples ($\sim90\%$) so 
we will not apply any corrections for the objects not detected in $H$. 
\citet[][]{xu06} and \citet[][]{igle06} have discussed the relative 
distribution of NIR luminosities (i.e. stellar masses) within former FIR and 
FUV-selected samples. 
Similar trends are found with the new samples but  since it is not the topic 
of the present work, we defer to these papers for a detailed discussion.

\begin{figure}
\centering
 \includegraphics [angle=-90,width=18cm]{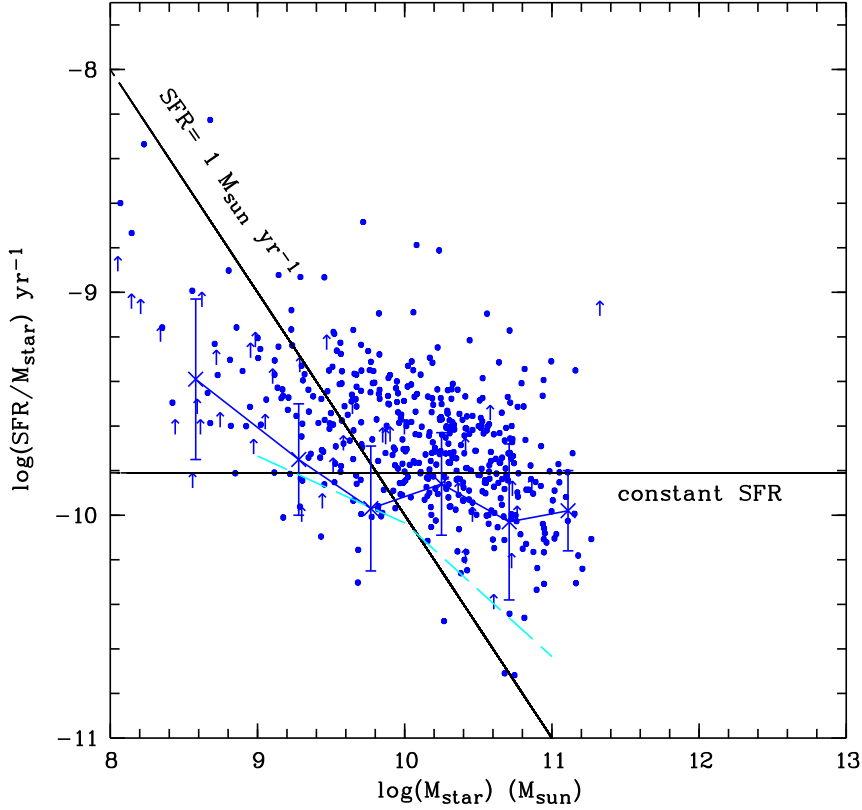}
   \caption{
     Specific star formation rate (SSFR; SFR per unit galaxy stellar mass) for the FUV-selected sample. 
     Blue solid line: average SSFR, 1 $\sigma$ errors are 
     overplotted. 
     The  horizontal dot-dashed line represents a  constant SFR over 
     the lifetime of the galaxy. 
     The diagonal line corresponds to a present SFR equal to 
     $1 M_\sun \mbox{yr}^{-1}$. 
     The average SSFR found by \citet[][]{Brinchmann} is plotted as 
     a cyan long dashed line
}
      \label{specific_sfruv}
   \end{figure}

\begin{figure}
\centering
 \includegraphics [angle=-90,width=18cm]{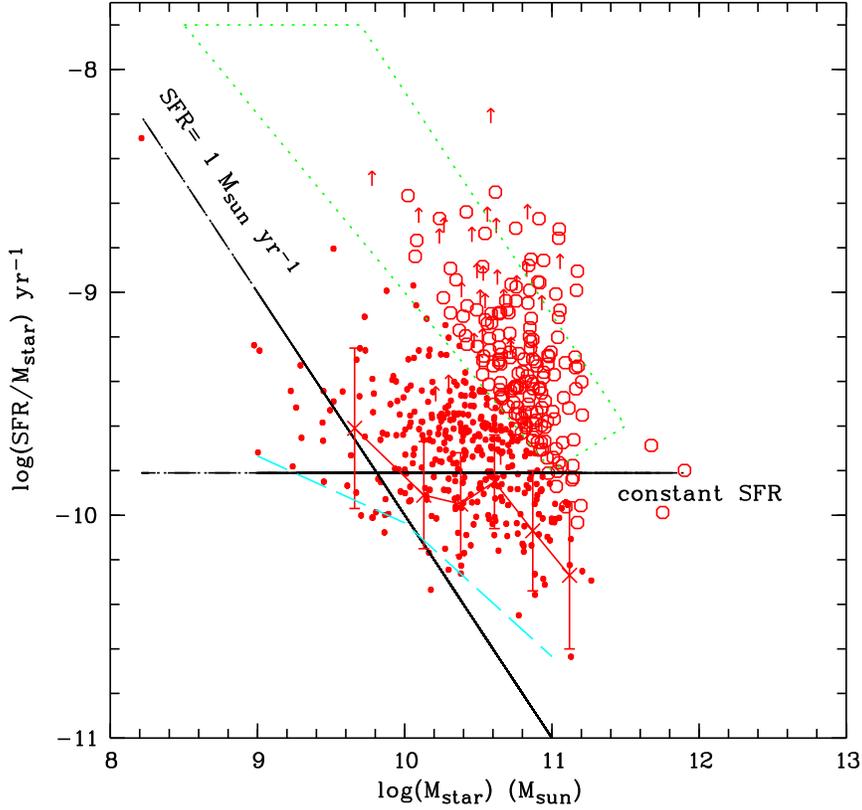}
   \caption{
     Specific star formation rate (SSFR)for the FIR-selected sample. 
     The empty circles represent the galaxies with $L_{\rm TIR}>10^{11} L_\sun$. 
     Red solid line: average specific SFR (SSFR), 1 $\sigma$ errors are 
     overplotted. 
     The  horizontal  line represents a  constant SFR over 
     the lifetime of the galaxy. 
     The diagonal  line corresponds to a present SFR equal to 
     1 Msun/yr. 
     The average SSFR found by \citet[][]{Brinchmann} is plotted as a cyan 
     long dashed line. 
     The dashed (green) box is the locus of the  galaxies  
     selected by \citet[][]{bell} at $z=0.7$.
}
      \label{specific_sfrir}
   \end{figure}

The  SSFR distributions are shown as a function of the stellar mass  
in Fig.~\ref{specific_sfruv} and Fig.~\ref{specific_sfrir}. 
Once again we must account for the very different volumes sampled by our 
two selections and so we accordingly calculate volume-weighted distributions to obtain 
the average trend of the specific SFR in the local universe.   
We define  
\begin{eqnarray}
  \langle \mbox{SSFR} \rangle \equiv 
    \frac{\sum_i \omega_i ~\mbox{SSFR}_i}{\sum{\omega_i}}
\end{eqnarray}
with  $\omega_i = 1/V_{\rm max}$ for each galaxy. 
The calculations are all performed with our Scenario 1 
(upper limits treated as detections) and we calculate geometrical means 
and standard deviations as in section 4.3 in order to be consistent 
with the logarithmic scales used in the study. 

In Fig.~\ref{specific_sfruv} are presented the results of the study for 
the FUV-selected sample. 
The SSFR is found to decrease  as the 
galaxy mass increases,   with and without applying 
a volume average. 
This confirms the trends found previously using  optically-selected 
samples at low and high redshifts \citep[e.g.][and references below]{cowie,boselli,heavens,bauer,feulner,panter}. 

We can compare our results more precisely with similar studies of the
SSFR at low redshifts.  \citet[][]{Brinchmann} performed a very
similar analysis (which inspired our own study) based on SDSS
spectroscopic data.  Their result is overplotted in
Fig.~\ref{specific_sfruv} where the birthrate parameter they
calculated is translated in SSFR as given in their paper (R=0.5
and t=13.7 Gyr in eq. (4)).  Similar trends are found; however, we
obtain larger SSFR for our galaxies selected in FUV and more massive
than $\sim 10^{10} M_\sun$.  \citet[][]{bauer} studied the evolution
of the SSFR-M$_\sun$ relation with redshift from spectroscopic surveys
of $K$-band and $I$-band selected samples and using the [OII]
$\lambda3727$ line to measure the SFR. We can compare our results to
theirs obtained for their lowest bins in redshift (z$<$0.25): almost
all their galaxies exhibit a SSFR distribution below the diagonal line which
corresponds to a SFR of $1\;M_\sun~ yr^{-1}$. Once again, most of
our galaxies, especially with $M> 10^{10} M_\sun$ exibit a SFR
larger than $1 \;M_\sun~ yr^{-1}$.  The spectroscopic data might be affected by
aperture effects although \citet[][]{Brinchmann} developed a
sophisticated method to compensate for this known effect. If we compare our
results with studies based on photometric data (i.e. without aperture
effects) and star formation rates deduced from the UV (rest-frame)
band like that of \citet[][]{feulner}, their mean value of the SSFR
found at z$\sim$0.6 and for massive galaxies ($10.5<log(M/M_\sun)<
11.5$) is consistent with our results, whereas the SSFR they found for
less massive galaxies at this redshift is higher than ours.

Therefore, our average values of SSFR for massive FUV-selected
galaxies appear larger than previously found from optical surveys at
low redshifts. Several reasons might be invoked to explain this
discrepancy: (i) as already suggested, the spectroscopic surveys may
suffer from aperture effects for nearby massive (and hence large angular sized)
galaxies; (ii) our FUV selection might be more biased toward galaxies
forming stars actively than the optical-NIR selections; (iii) our
study accounts for the stellar emission re-processed by dust in a
direct way whereas the dust attenuation in other studies is deduced
from UV-optical data only. The dust attenuation is known to increase
with the bolometric luminosity and so does the mass of those
galaxies. This parameter therefore may play a major role for the
highest mass bins; (iv) our derivation of the SFR assumes a correction
for the dust heating by the old stars (eq.(5)). A variation of $\eta$
with the stellar mass may affect our results, however to obtain values
of the SSFR as low as those found by \citet[][]{Brinchmann} we would have to
take $\eta=0.7$ instead of 0.3, which seems very unlikely for a galaxy
sample dominated by intermediate types for the high mass range (Sbc-Sc
galaxies).

The situation appears even more complex for the FIR-selected sample as shown 
in Fig.~\ref {specific_sfrir}. The sample is shifted toward more massive galaxies 
as compared to the FUV-selected sample \citep[e.g.][]{xu06}. It also contains a large 
fraction of bright 
galaxies  ($L_{\rm TIR}>10^{11} L_\sun$) which do not follow the general trend found in FUV and optical selections: these very bright and massive galaxies exhibit a high  SSFR. 
However these galaxies are rare objects in the  nearby universe and as 
soon as a volume average is performed we also find a global decrease of 
the mean SSFR when the mass increases for the FIR selection. Thus the decrease of the SSFR as the galaxy mass increases  
appears to be independent of the sample selection in 
the nearby universe (FIR, FUV or optical) and reflects an intrinsic mean 
property of the local universe. The average curve is consistent with that found for 
the FUV selection if we account for error bars (at the one-$\sigma$ level), but it is above the 
relations found by \citet{Brinchmann} and \citet[][]{bauer} at low z.

Finally, it is of some interest to compare our results with studies  which 
also include FIR data. \citet[][]{caputi} have compared the stellar masses 
and the star formation rates of galaxies selected at 24 $\mu$m from z = 0.5 to 3. 
If we extrapolate the trend they found down to z=0 (in their Fig. 10) our average 
values of the SSFR at z=0 are consistent with their results.
\citet[][]{bell} analysed a rest-frame $B$-band selected sample of galaxies 
at $z=0.7$ and cross-correlated it with data at 24 $\mu$m 
(detection rate 1/3). 
The locus of the galaxies they detected at 24 $\mu$m is shown in 
Fig.~\ref{specific_sfrir}. 
Their sample is complete only for galaxies more massive than 
$2 ~ 10^{10} M_\star$. 
It is clear that these galaxies are more active than the average at $z=0$ 
found either in the FUV or the FIR selection, thus confirming the 
conclusions of \citet[][]{bell}. 
The massive galaxies detected at $z=0.7$ have a SSFR similar to what is 
seen in nearby LIRGs although their morphology may well be different 
\citep[][]{melbourne}.

\section{Conclusions}

\begin{enumerate}
\item We have built large FIR and FUV-selected galaxy samples fully 
representative of the nearby universe. 
The two selections  are found to sample very different volumes, which 
is a direct consequence of the very different shape of the 
luminosity functions at the two wavelengths. 
Therefore one  must apply volume corrections before comparing 
the mean properties of these samples. 

\item The bolometric luminosity of newly formed stars is estimated by 
combining the infrared and ultraviolet fluxes and accounting for dust heating by 
old stars.  The bolometric  LF is calculated and  found to be different in the FUV and 
the FIR selections. Intrinsically bright galaxies are under-sampled by a FUV selection. 
No faint galaxy  ($L_{\rm bol} < 10^9 L_\sun$) is found  in our FIR sample. We must 
wait for deeper FIR imaging surveys with a better spatial resolution than {\sl IRAS} to 
compare accurately the lowest bins of the  bolometric luminosity functions and investigate 
the contribution of close pairs and interacting systems. 

\item The ratio of the total IR luminosity to the FUV  
($L_{\rm TIR}/L_{\rm FUV}$) is found to be strongly related to  the 
TIR and to the bolometric luminosity for both samples. 
No universal trend is found with the FUV luminosity, making irrelevant any 
dust attenuation correction based on a monochromatic FUV luminosity alone. 
The volume-averaged relation between $L_{\rm bol}$ and 
$L_{\rm TIR}/L_{\rm FUV}$ is found to be similar for both (FUV and FIR-selected) 
samples for bolometric luminosities between $10^9 L_\sun$ to 
$5~10^{10} L_\sun$. 
The monotonic increase of $L_{\rm TIR}/L_{\rm FUV}$ with 
$L_{\rm bol}$ continues up to $10^{12} L_\sun$ within the FIR selection. 
$L_{\rm TIR}/L_{\rm FUV}$ saturates for FUV-selected galaxies 
more luminous than $5 ~ 10^{10} L_\sun$ at a value corresponding to a dust attenuation of $\sim 2.5$ mag in FUV. 

\item The specific star formation rate is analysed as a function of 
the stellar mass. 
It is found to decrease as the galaxy mass increases at both wavelengths and as soon as volume corrections are applied to the samples. Massive, FUV-selected galaxies and all the FIR-selected ones exhibit a larger 
specific star formation rate than that deduced from optical-NIR surveys 
of nearby galaxies with similar stellar masses selected in optical or in the NIR. 
Luminous FIR-selected galaxies ($L_{\rm bol}>10^{11}~L_\sun$)  have 
a very large specific star formation rates,  similar to those found at 
$z=0.7$ by \citet[][]{bell} for FIR-luminous galaxies of similar mass and luminosities.
\end{enumerate}

\acknowledgments
We thank the anonymous referee for her/his very 
useful and extensive comments.
{\sl GALEX} (Galaxy Evolution Explorer) is a NASA Small Explorer, launched in April 2003.
We gratefully acknowledge NASA's support for construction, operation,
and science analysis for the {\sl GALEX} mission,
developed in cooperation with the Centre National d'Etudes Spatiales (CNES)
of France and the Korean Ministry of 
Science and Technology. 
V.B, D.B. and J.I-P. gratefully acknowledge CNES and ''Programme National Galaxie'' support for science analysis for 
the {\sl GALEX} mission. 
TTT has been supported by JSPS Fellowship for Research Abroad for the 
early phase of this project, and later by 
The 21st Century Center-of-Excellence Program
``Exploring New Science by Bridging Particle-Matter Hierarchy'', Tohoku
University.
This research has made use of the NASA/IPAC Extragalactic Database (NED) which is operated by the Jet Propulsion Laboratory, California Institute of Technology, under contract with the National Aeronautics and Space Administration. We also acknowledge the usage of the HyperLeda database (http://leda.univ-lyon1.fr).

\clearpage

\begin{table}
\caption{FIR-selected sample:(1) {\sl IRAS} name from the {\sl IRAS} Point Source Catalog (PSC), (2) flux at 60 $\mu$m in Jansky, (3) Flux at 100 $\mu$m in Jansky, (4) FUV magnitude (AB scale) corrected for Galactic extinction (see text), when no value is quoted an upper limit ar 20.5 mag is adopted (see text), (5) radial velocity in km s$^{-1}$ from the PSC$z$, (6) total H magnitude from 2MASS, (7) morphological type from HyperLeda or NED  (the values taken from NED are preceded by ".")}             
\label{table:1}      
\centering          
\begin{tabular}{c c c c c c c}     
\hline\hline       
{\sl IRAS} name & $F_{60}$ & $F_{100}$ & FUV & cz & H & Type\\ 
PSC& Jansky&Jansky&mag&km s$^{-1}$&mag&HyperLeda/NED\\
\hline        
    IRAS00003-0747 & 0.65 &  0.69 & 19.75&8697 & 12.72 & Sab\\     
    IRAS00005-0211 & 0.84 &  2.58 &   16.60 & 7271& 10.53 & Sbc\\  
    IRAS00007+0235 & 1.08 &  1.38 &  *&27855 &13.82& *\\             
    IRAS00022-0150 & 0.96 &  1.41 & 16.65&7134 & 11.94&   Sb\\ 
    IRAS00025-0722 & 1.07 &  3.22 & 15.57 & 3765 &10.53 & SBb\\
\hline                  
\end{tabular}
\end{table}
\begin{table}
\caption{FUV-selected sample:(1) PGC (from HyperLeda)  or 2MASS number (only 3 objects have only a 2MASS number, they are at the end of the table), (2) FUV magnitude (AB scale) corrected for Galactic extinction (see text), (3) flux at 60 $\mu$m in Jansky when no value is quoted and upper limit of 0.2 Jy is adopted, (4) Flux at 100 $\mu$m in Jansky, when no value is quoted a mean ratio for $F_{60}/F_{100}$ is adopted (see text)(5) radial velocity in km s$^{-1}$ from HyperLeda and NED by order of preference (the values taken from NED are indicated by a "N", (6) total H magnitude from 2MASS, (7) morphological type from HyperLeda and NED (the values taken from NED are preceded by ".")}             
\label{table:2}      
\centering          
\begin{tabular}{c c c c c c c}     
\hline\hline       
Galaxy name & FUV & $F_{60}$ & $F_{100}$ & cz & H & Type\\ 
PGC/2MASS& mag&Jansky&Jansky&km s$^{-1}$&mag&HyperLeda/NED\\
\hline        
PGC 158 & 16.87 &  1.65  & 3.15 & 19199& 12.02   &  I\\
PGC  229 & 16.82 &  0.65  & 2.10  &6202 & 12.41 & SABc\\
PGC   282 & 16.43&   0.28 &  0.66 & 11406& 12.76 & Sa\\
PGC 305 & 15.49 &  * &  * &3112 &  12.75  & Sc\\              
PGC  312  &15.50  & 1.04  & 3.27 & 3816& 10.53 &  SBb \\

\hline                  
\end{tabular}
\end{table}

\clearpage


\end{document}